\documentclass{article}

\usepackage{arxiv}

\usepackage[utf8]{inputenc} 
\usepackage[T1]{fontenc}    
\usepackage{hyperref}       
\usepackage{url}            
\usepackage{booktabs}       
\usepackage{amsfonts}       
\usepackage{nicefrac}       
\usepackage{microtype}      
\usepackage{lipsum}
\usepackage{graphicx}
\usepackage{graphics}
\graphicspath{ {./images/} }
\usepackage{bm}
\usepackage{amsthm}
\usepackage{mathtools, nccmath}
\usepackage{amsmath,amsfonts,amssymb,amsthm}
\usepackage{mathtools}
\usepackage{lipsum}
\usepackage{siunitx,booktabs,cuted,capt-of}
\usepackage{tabularx,ragged2e}
\usepackage{multirow}

\newcolumntype{C}{>{\Centering\arraybackslash}X}

\title{Defocus Deblur Microscopy via head-to-tail cross-scale fusion}

\name{Jiahe Wang$^{\dagger}$ \qquad Boran Han$^{\ddagger \ast}$ \thanks{$^\ast$Corresponding author.}}
  
\address{$^{\dagger}$ School of Computing and Information, University of Wisconsin-Madison, Madison, WI 53715 \\
$^{\ddagger}$ Shell International Exploration and Production Inc, Boston, MA 02210}

\author{
 Jiahe Wang \\
  School of Arts and Sciences\\
  University of Wisconsin-Madison\\
  Madison, WI 53706 \\
  \texttt{jwang2249@wisc.edu} \\
   \And
 Boran Han \\
Shell International Exploration and Production Inc, MA 02210 \\
  \texttt{ZIL50@pitt.edu} \\
  \And
}

\begin{document}
\maketitle
\begin{abstract}

Microscopy imaging is vital in biology research and diagnosis. When imaging at the scale of cell or molecule level, mechanical drift on the axial axis can be difficult to correct. Although multi-scale networks have been developed for deblurring, those cascade residual learning approaches fail to accurately capture the end-to-end non-linearity of deconvolution, a relation between in-focus images and their out-of-focus counterparts in microscopy. In our model, we adopt a structure of multi-scale U-Net without cascade residual leaning. Additionally, in contrast to the conventional coarse-to-fine model, our model strengthens the cross-scale interaction by fusing the features from the coarser sub-networks with the finer ones in a head-to-tail manner: the decoder from the coarser scale is fused with the encoder of the finer ones. Such interaction contributes to better feature learning as fusion happens across decoder and encoder at all scales. Numerous experiments demonstrate that our method yields better performance when compared with other existing models.

\end{abstract}


\section{Introduction}

The quality of images taken by microscopes is imperative to biology-related research as well as disease diagnoses, such as malaria and cancer detection. However, keeping the biology sample of the target in focus can be challenging due to the imperfect microscopy's mechanical stability. The drift caused by instability is called mechanical drift \cite{Carter:07}. While lateral drift (x and y direction) can be corrected by image registration \cite{articleWang}, drift along the axial axis (z-direction) poses a significant challenge in correction. If not corrected, such drift can decrease the resolution of microscopic imaging. In the case of microscopes with a high-precision objective lens, samples taken at 1 $\mu$ m away from the focal plane can lead to tremendous image quality degradation. However, such a high-precision objective lens can be essential for single-molecule tracking and molecule biology research. 

A conventional solution includes improving mechanical stability \cite{ahmad2020highly} or installing a real-time feedback system, such as focus lock \cite{Carter:07}. One prevalent method is to make an infrared beam reflected off a cover-slip and calculate the shifts in the position of this beam to maintain constant objective-slide separation by closed-loop control. However, such a method is also found to be vulnerable when correcting significant drift. In addition, out-of-focus (OOF) images can be wasted if mechanical drift can't be corrected since such an inconvenient issue will lead to data re-acquisition.

\begin{figure}[h]
	\centering
	\includegraphics[width = 0.9\linewidth]{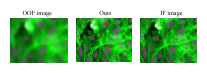}
\caption{Example OOF image (left), predicted IF images from our method (middle) and ground truth IF images (right) }
	\label{model_figure}
\end{figure}

An alternative solution to avoid data re-acquisition is to deblur the OOF image using image processing methods. Traditionally, deconvolution has been implemented to recover the in-focus (IF) images from OOF images \cite{Sibarita2005}. This method usually requires to have point spread function (PSF) \cite{Sibarita2005}. However, in practice, finding the true PSF is impossible. Therefore, an approximation of PSF is usually obtained by theoretical calculation based on the numerical aperture (N. A.) or measurement from some known probes (i. e., beads). Recently, defocus deblurring using deep learning has become a popular topic thanks to convolution neural networks (CNN). Such methods are most commonly used to process blurred pictures taken by cameras \cite{Lee_2021_CVPR, Lee_2019_CVPR}. In contrast to the conventional imaging system, i.e., camera, the microscopic image is often required to be accurate. Despite the fact that limited methods have been introduced focuses on solving the deblurring problems for microscopy images \cite{https://doi.org/10.1002/jbio.201960147, article, Liu:20, NA2021116987, DBLP:journals/corr/abs-2011-11879}, some of the methods are tested under synthetic defocus data \cite{DBLP:journals/corr/abs-2011-11879, https://doi.org/10.1002/jbio.201960147}, while others either fail to provide an end-to-end method \cite{article}

\begin{figure*}[h]
	\centering
	\includegraphics[width=0.8\textwidth]{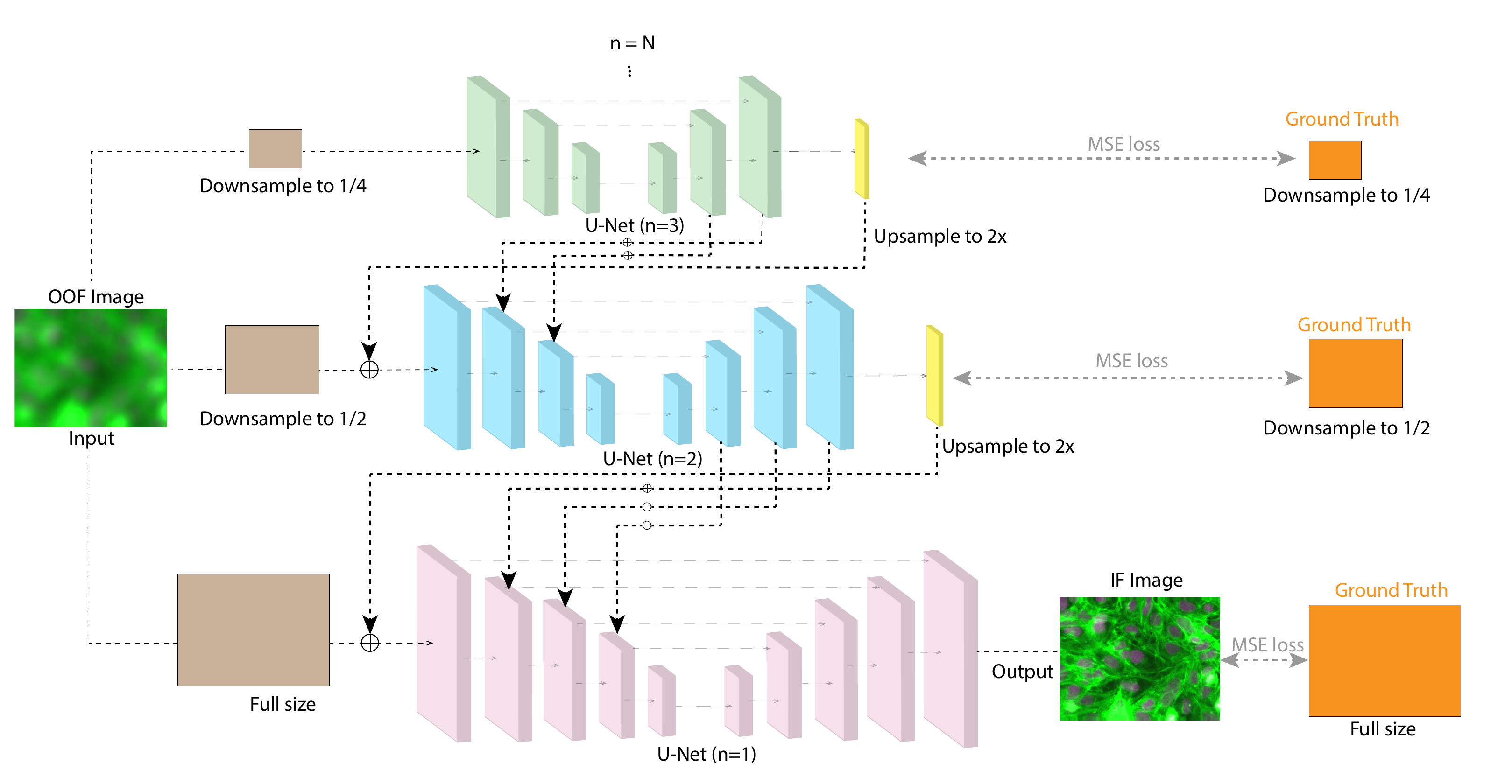} 
\caption{The structure of our model. OOF images will be down-sampled and fed into corresponding multi-scale U-Net. The hidden features from the decoder of coarser sub-networks will be fused to the encoder of the finer sub-networks (marked as symbol $\bigoplus$ ), improving the training efficiency.}
	\label{model_figure}
\end{figure*}

To overcome these limitations, we develop an end-to-end deep learning-based workflow to predict IF microscopic images from OOF counterparts. Inspired by  \cite{Nah_2017_CVPR}, we adopt a multi-scale U-Net, with feature fusion across scales in a head-to-tail manner. We test two fusion modes and demonstrate that our proposed architecture performs better than other competing methods. 

\section{Method}

\subsection{Prior knowledge}

\textbf{Problem defined.} An OOF image ($x$) can be obtained from IF image ($y$) convoluted by the PSF at a given $z$:
\begin{equation}
\label{convolution}
x = y*PSF(z) + \epsilon
\end{equation}
where $*$ is the convolution operator and $\epsilon$ is the noise. To obtain $y$ from $x$, an inverse operator of convolution is needed, i. e. deconvolution.  

\textbf{Coarse-to-fine network.} As a pioneering work,  \cite{Nah_2017_CVPR} directly learns the relation between blurry-sharp image pairs in an end-to-end manner by adopting a coarse-to-fine strategy. Each subnetwork consists of a sequence of convolutional layers that maintains the spatial resolution of input feature maps. Different scales of input images are fed into the corresponding sub-networks. The resultant image from a coarser scale sub-network is added to the input of a finer scale sub-network to enable coarse-to-fine information transfer. The reconstruction procedure of \cite{Nah_2017_CVPR} is formulated as follows \cite{https://doi.org/10.48550/arxiv.2108.05054}:

\begin{equation}
\label{finetocoarse}
\hat{y}_n = U^{(n)}_{\theta_n}(x_n;\hat{y}_{n+1} \uparrow) + x_n
\end{equation}
where $U^{(n)}_{\theta_n}$ is the $n$-th subnetwork parameterized by $\theta_n$. $\hat{x}_n$ and $\hat{y}_n$ are blurred and predicted deblurred images at $n$-th scale. $\uparrow$ is the upsampling operation. It is based on the assumption that the deblurred image can be obtained by adding the fine details to the blurred images (Eq. \ref{finetocoarse}). 

However, such an assumption is inaccurate in the microscopy system, as such scheme adopts a linear summation of iterative nonlinear correction, following the minor perturbation manner. However, the inverse operation of Eq. \ref{convolution} (deconvolution) is the end-to-end nonlinear relation between $x$ and $y$. Accordingly, in the following section, we proposed a model $M: x \rightarrow y$ that captures the non-linearity of deconvolution while preserving the coarse-to-fine multi-scale strategy.

\subsection{Defocus debluring model}

\begin{table*} [t]
\begin{center}
\begin{tabular}{cccccccc} 
 \hline
Distance & & 34 $\mu$m & 28$\mu$m & 22$\mu$m & 16$\mu$m & 10$\mu$m & No. of trainable parameters\\ 
  \hline
    \hline
$N = 1$ (U-Net) & PSNR & 34.013 & 34.868 & 36.182 & 36.455 & 38.409 & 1,942,962\\ 
 & SSIM & 0.8482 & 0.8737 & 0.9013 & 0.9260 & 0.9561 & \\ 
  \hline
  \hline
  Summation fusion &&&&&&\\
    \hline
$N = 2$ & PSNR & 33.608 & 35.122 & 35.582 & 37.399 & 38.991 & 1,986,116\\ 
 & SSIM & 0.8451 & 0.8753 & 0.8989 & 0.9295 & 0.9591 & \\ 
  \hline
 $N = 3$ & PSNR & 34.641 & 35.619 & 36.418 & 37.474 & 39.831 &2,543,478\\ 
  & SSIM & 0.8454 & 0.8706 & 0.8990 & 0.9263 & 0.9576 & \\ 
   \hline
   $N = 4$ & PSNR & 34.552 &  \textbf{35.688}& 36.352  & \textbf{37.993} & \textbf{40.860} &  2,569,512 \\ 
    & SSIM & \textbf{0.8491} & \textbf{0.8721} & 0.8978 & \textbf{0.9309} & \textbf{0.9612} & \\ 
  \hline
    \hline
    Concatenation fusion &&&&&&\\
    \hline
$N = 2$ & PSNR  & 34.212  & 35.086 & 34.361 & 37.407 & 38.932 &1,986,116\\ 
 & SSIM & 0.8462 & 0.8708 & 0.8907 & 0.9265 & 0.9529 & \\ 
  \hline
$N = 3$ & PSNR  & 33.928 & 35.402 & 36.298 & 37.914 & 39.725 & 2,543,478 \\ 
 & SSIM & 0.8452 & 0.8702 & 0.8983 & 0.9296 & 0.9589 & \\ 
  \hline
$N = 4$ & PSNR  &\textbf{34.782} & 35.596  & \textbf{36.448}  & 37.591 & 40.583 & 2,942,760 \\ 
 & SSIM & 0.8487 & 0.8708 & \textbf{0.8996} & 0.9299 & 0.9602 & \\ 
 \hline
\end{tabular}
\caption{PSNR and SSIM values of our model with different number of scales ($N$) and feature fusion modes. Best in bold.}
\end{center}
\vspace{-2mm}
\end{table*}

\begin{table*} [t]
\centering
\begin{tabular}{cccccccc} 
 \hline
Distance & Metric & 34 $\mu$m & 28$\mu$m & 22$\mu$m & 16$\mu$m & 10$\mu$m & No. of trainable parameters\\ 
  \hline
   Input& PSNR & 32.048 & 32.421 & 32.915 & 33.732 & 35.404 & N/A \\ 
    & SSIM & 0.7161 & 0.7572 & 0.7861 & 0.8300 & 0.8954 \\ 
  \hline
 Nah et. al. \cite{Nah_2017_CVPR} & PSNR & 32.810  & 33.705 &  35.172 & 36.419 & 37.763 & 1,772,740 \\ 
   & SSIM & 0.8337 & 0.8616 & 0.8934 & 0.9253 & 0.9552 \\ 
  \hline
 Liu et. al. \cite{Liu:20} &  PSNR & 31.011 & 32.872 &  34.030 & 35.098 & 37.199 & 1,461,124 \\ 
   & SSIM & 0.7171 & 0.7704 & 0.8442 & 0.9051 & 0.9469 \\ 
   \hline
 Na et. al. \cite{NA2021116987} & PSNR & 31.409  & 31.908 &  32.513 & 34.804 & 37.253 & 4,384,262\\
   & SSIM & 0.7283 & 0.7586 & 0.8074 & 0.8908 & 0.9407\\ 
   \hline
  Zhao et. al. \cite{https://doi.org/10.1002/jbio.201960147} & PSNR & 34.125 & 35.153 & 36.412  & \textbf{37.995} &  40.420 & 811,778\\ 
  & SSIM & 0.8260 & 0.8643 & 0.8952  & 0.9285 & \textbf{0.9612} \\ 
   \hline
 Ours (Summation fusion) & PSNR & 34.552 & \textbf{35.688}& 36.352  & 37.993 & \textbf{40.860} & 2,569,512 \\
 & SSIM & \textbf{0.8491} & \textbf{0.8721} & 0.8978 & \textbf{0.9309} & \textbf{0.9612} \\ 
  \hline
 Ours (Concatenation fusion) & PSNR & \textbf{34.782} & 35.596  & \textbf{36.448}  & 37.591 & 40.583  & 2,942,760\\ 
 & SSIM & 0.8487 & 0.8708 & \textbf{0.8996} & 0.9299 & 0.9602 \\ 
 \hline
\end{tabular}
\caption{PSNR and SSIM values of our model and other existing models. Best in bold.}
\vspace{-0mm}
\end{table*}

We exploit U-Net \cite{DBLP:journals/corr/RonnebergerFB15} of each sub-network\cite{Gao_2019_CVPR}. However, instead of learning the finer residual of the images and adding to the blurred images, we fuse distilled information between hidden layers between the adjacent sub-networks. Such a feature fusion has been shown to boost the representation power of CNNs, such as Residual Networks (ResNet) \cite{7780459}. In our model, the output of $n$-th sub-network is obtained from blurred images $x_n$ and the output of $n+1$-th sub-network, $\hat{y}_{n+1}$. $n$-th sub-network ($U^{(n)}_{(\theta_n, \theta_{n+1})}$) is parameterized by $\theta_n$ and $\theta_{n+1}$ when $n<N$:

\begin{equation}
\label{model}
\hat{y}_n = U^{(n)}_{(\theta_n, \theta_{n+1})}(x_n;\hat{y}_{n+1} \uparrow)
\end{equation}
and is parameterized by $\theta_n$ when $n=N$

\begin{equation}
\label{model}
\hat{y}_n = U^{(n)}_{(\theta_n)}(x_n)
\end{equation}
where $N$ is the number of scales, or subnetworks. $U^{(n)}$ denotes the encoder-decoder-based U-Net with symmetric skip connections that directly transfers the feature maps from the encoder to the decoder. In contrast to conventional U-Net, the encoder $U^{(n)}$ of our model is also connected with the decoder of $U^{(n+1)}$, shown in Figure \ref{model_figure}. Intuitively, our finer model can enable more efficient learning via utilizing the knowledge distilled from coarser counterparts. In detail, the $U^{(n)}$ can be represented as $(E^{(n)}_0, E^{(n)}_1, \cdots, E^{(n)}_i, \cdots, E^{(n)}_I)$ and $(D^{(n)}_0, D^{(n)}_1, \cdots, D^{(n)}_i, \cdots, D^{(n)}_I)$, where $E^{(n)}_i$ and $D^{(n)}_i$ are $i$-th features maps in the encoder and decoder of $U^{(n)}$, respectively. As a result, $E^{(n)}_i$ can be obtained by the following equation:
\begin{align}
\label{depth classifier model}
E^{(n)}_i &= F(E^{(n)}_i, D^{(n+1)}_{i-1})
\end{align}
$F$ is the feature fusion operator which takes the $E^{(n)}_i$ and $D^{(n+1)}_{i-1}$. 

We use the multi-scale loss function, defined as follows:
\begin{equation}
\label{depth classifier model}
L = \sum^N_{n=1} (\hat{y}_n-y_n)^2
\end{equation}
 where $y_n$ is the down-sampled ground-truth. By this mean, the output from each scale is constrained by the ground truth, ensuring that the features are correctly trained before fusion.

\subsection{Directional Feature Fusion}

We here apply two methods to fuse hidden features: summation and concatenation. In residual learning mode, the learnt features from decoder layer in $n+1$ subnetwork ($ D^{(n+1)}_{i-1}$) are added to encoder layer of $n$ subnetwork ($E^{(n)}_i$):

\begin{align}
\label{depth classifier model}
E^{(n)}_i &= E^{(n)}_i + D^{(n+1)}_{i-1}
\end{align}

Alternatively, $E^{(n)}_i$ can also be obtained by channel aware concatenating the $ D^{(n+1)}_{i-1}$ :

\begin{align}
\label{concat_mode}
E^{(n)}_i = [E^{(n)}_i : D^{(n+1)}_{i-1}]
\end{align}
where $:$ is the concatenation operator. The above equation (Equation \ref{concat_mode}) depicts feature concatenation mode. Our proposed fusion is directional: decoder features from coarser scale sub-network are fused to the encoder from finer scale sub-network.

\section{Experiments}

\subsection{Experiment Settings}

\textbf{Dataset.} Our data sets come from the Broad Bioimage Benchmark Collection \cite{articleLjosa} (https://bbbc.broadinstitute.org/BBBC006). The data sets consist of 34 sets of microscopic images from human U2OS cells. The cells are stained with Hoechst 33342 markers and actin markers by phalloidin.

The stained cells are imaged with an exposure of 15 and 1000 ms for two colors sequentially using a 20x objective. 
For each site, the optimal focus is found via laser auto-focusing. The automated microscope was then programmed to collect a z-stack of 32 image sets with the step size of 2 $\mu$m. As instructed by the dataset, we set z = 17 as our ground truth. Each z position consists of 768 images from different imaging areas. Each image, containing two-color channels, has a size of 696 x 520 pixels in 16-bit TIF format. In our experiment, we assign the first 675 pictures as the training set and the rest 93 images as the test set. We segment each image into four images with the size of 348 x 260 for ease of computation. 

\textbf{Training and evaluation} We use an Adam optimizer with a learning rate of 0.01, a momentum of 0.9, and a batch size of 8. The network is trained for 80 epochs. Because the PSF is depth-dependent, our model is trained from scratch for every depth for evaluation. For a fair comparison, all other competing methods are trained following the same process. SSIM and PSNR \cite{1284395, Hor2010ImageQM} are measured for performance estimation. The following experiments were performed on a desktop with NVIDIA RTX 3080 GPU.

\subsection{Experiment Results}

\textbf{Ablation study.} We evaluate our model using a various number of sub-networks ($N = 1-4$) and notice that performance improves as $N$ increases. The result of the 2-level U-Net performs only slightly better than the original U-Net model. However, when we added the third and fourth levels of the U-Net, the model's performance increased substantially. This result could be proven by the PSNR and SSIM values shown in Table 1. We also assess the two feature fusion modes proposed in Section 2.3. Our results demonstrate that the summation mode performs slightly better than the concatenation mode. Additionally, the summation mode contains less trainable parameters. 

\textbf{Performance against other methods.} To demonstrate our approach is able to accurately predict IF images, we compare our method with other existing methods, including methods from Nah et al. \cite{Nah_2017_CVPR}, Liu et al. \cite{Liu:20}, Na et al. \cite{NA2021116987} and Zhao et al. \cite{https://doi.org/10.1002/jbio.201960147}. Among them, the study from Nah et al. \cite{Nah_2017_CVPR} has been demonstrated using camera images, such as GoPro and Kohler Dataset. At the same time, the rest are specifically designed for the optical or electron microscopic imaging domain. We also note that not all methods provide source codes. Therefore, we implement their models based on the information provided in the original papers. Liu et al. \cite{Liu:20} use residual connections for denoising and deblurring; Na et al. \cite{NA2021116987} employ the multi-scale network consisting of several ResNet \cite{7780459}; Zhao et al. \cite{https://doi.org/10.1002/jbio.201960147} adopt the residual DenseNet (RDN) \cite{zhang2018residual}, which is commonly used for super-resolution imaging. Table 2 shows that our proposed method ($N=4$) can predict the IF images more accurately. Additionally, we measured the SSIM and PSNR between the input and the label, serving as the baseline. The increase beyond baseline demonstrates the capability of image restoration using deep learning models. 

\section{Conclusion and Discussion}

This paper develops a new coarse-to-fine network with head-to-tail feature fusion learning for microscopic image defocus deblurring. Our method shows superior results when tested with fluorescent images. 

\textbf{Advantages compared to coarse-to-fine multi-scale network \cite{Nah_2017_CVPR} and other feature fusion-based modified models \cite{dong2020multiscale, LI2021103149}.} Our model, describes in Eq. \ref{model}, depicts an end-to-end non-linear function between $x$ and $y$ via coarse-to-fine hidden feature learning ($E^{(n)}_i$). 
However, most of the existing networks \cite{Nah_2017_CVPR, dong2020multiscale} learn the final output through a residual cascade manner ($\Delta y$). Another difference stems from our head-to-tail interaction between adjacent sub-networks. This unique interaction enables feature fusion between the decoding features of the coarser scale sub-network and encoding features of the finer scale sub-network. The fused features can be efficiently learned through another encoder-decoder within the same scale sub-network, which can be advantageous compared with conventional cross-layer feature fusion \cite{LI2021103149}. 

For future study, our head-to-tail interaction can also be potentially combined with attention gates. Additionally, we note that the performance of our method in other imaging domains awaits further evaluation. 

\bibliographystyle{IEEEbib}  

\bibliography{main}

\end{document}